\documentclass[12pt,letterpaper]{article}
\usepackage{amscd}

\global\def\draftcontrol{0}

   \def\versionno{ Mesons in WQCD }
\catcode`\@=11

\expandafter\ifx\csname draftcontrol\endcsname\relax\global\def\draftcontrol{0}
\fi

{\count255=\time\divide\count255 by 60
\xdef\hourmin{\number\count255}
\multiply\count255 by-60\advance\count255 by\time
\xdef\hourmin{\hourmin:\ifnum\count255<10 0\fi\the\count255}}
\def\draftdate{\number\month/\number\day/\number\year\ \ \ \hourmin }

\newcommand\makepapertitle{\par
  \begingroup
    \renewcommand\thefootnote{\@fnsymbol\c@footnote}%
    \def\@makefnmark{\rlap{\@textsuperscript{\normalfont\@thefnmark}}}%
    \long\def\@makefntext##1{\parindent 1em\noindent
            \hb@xt@1.8em{%
                \hss\@textsuperscript{\normalfont\@thefnmark}}##1}%
     \newpage
     \global\@topnum\z@   
     \@makepapertitle
     \thispagestyle{empty}\@thanks
  \endgroup
  \setcounter{footnote}{0}%
  \global\let\thanks\relax
  \global\let\makepapertitle\relax
  \global\let\@makepapertitle\relax
  \global\let\@thanks\@empty
  \global\let\@author\@empty
  \global\let\@date\@empty
  \global\let\@title\@empty
  \global\let\title\relax
  \global\let\author\relax
  \global\let\date\relax
  \global\let\and\relax
  \def\version{\let\version\@version\@gobble}
}
\def\@makepapertitle{%
  \newpage
   \ifnum\draftcontrol=1 {}
   \version\versionno
   \vskip 3em%
   \else
   \hfill\hbox to 3cm {\parbox{4cm}{\@pubnum}\hss}%
   \vskip 3em%
   \fi
   \begin{center}%
   \let \footnote \thanks
     {\LARGE {\@title}}%
     \vskip 1.5em%
     {\normalsize
       \lineskip .5em%
       \begin{tabular}[t]{c}%
         \@author
       \end{tabular}\par}%
     \vskip 1.5em%
     {\@bstract}%
     \end{center}%
     \vskip 1.5em 
     \@date%
   \par
}

\gdef\@pubnum{}
\def\pubnum#1{%
  \gdef\@pubnum{#1}}

\gdef\@bstract{}
\def\Abstract#1{%
  \gdef\@bstract{%
   \parbox{\textwidth-0pc}{%
   \centerline{\bf Abstract}\penalty1000%
\noindent
\renewcommand\baselinestretch{1.0}%
{#1}}}
}

\def\ps@paper{\let\@mkboth\@gobbletwo%
     \ifnum\draftcontrol=1
        \def\@oddfoot{\hbox to \textwidth{\tiny \versionno \hfil\tiny\draftdate}%
        \hskip -\textwidth \hbox to \textwidth{\hfil\rm\thepage\hfil}}%
     \else\def\@oddfoot{\hbox to \textwidth{\hfil\rm\thepage\hfil}}
     \fi
     \let\@evenfoot\@oddfoot
}

\def\@version#1{\ifnum\draftcontrol=1
\typeout{}\typeout{#1}\typeout{}
\vskip3mm\centerline{\hbox{\fbox{\normalsize{\tt DRAFT -- #1 -- }
                   {\draftdate}}}}\vskip3mm
\fi}
\let\version\@version
\long\def\eqlabel#1{\ifnum\draftcontrol=1
                    \tag@false  
                    \tag*{(\theequation) \hbox to -0.2cm{\hspace{0cm}\small{#1}\hss}}
                    \refstepcounter{equation} 
                    \edef\@currentlabel{\theequation}
                    \ltx@label{#1}          
                    \else
                    \label{#1}
                    \fi
                    }
\let\st@bibitem\@bibitem
\let\st@lbibitem\@lbibitem
\ifnum\draftcontrol=1
  \def\@bibitem#1{%
    \st@bibitem{#1}\a@@label{#1}\ignorespaces}
  \def\@lbibitem[#1]#2{%
    \st@lbibitem[#1]{#2}\a@@label{#2}\ignorespaces}
  \def\a@@label#1{%
    \gdef\a@lab{\smash{\normalfont\small#1}}
    \ifvmode
      \if@inlabel
        \global\setbox\@labels\hbox{%
          \llap{\a@lab\let\a@lab\relax
                \kern\@totalleftmargin\kern\marginparsep}%
          \box\@labels}%
      \fi
    \fi}
\fi

\usepackage{amsmath,amssymb,array,calc,rotating,epsfig,psfrag}

\ifnum\draftcontrol=1
\fi

\renewcommand\baselinestretch{1.25}
\renewcommand\section{\@startsection {section}{1}{\z@}%
                                   {-3.5ex \@plus -1ex \@minus -.2ex}%
                                   {2.3ex \@plus.2ex}%
                                   {\normalfont\large\bfseries}}
\renewcommand\subsection{\@startsection{subsection}{2}{\z@}%
                                   {-3.25ex\@plus -1ex \@minus -.2ex}%
                                   {1.5ex \@plus .2ex}%
                                   {\normalfont\normalsize\bfseries}}
\renewcommand\subsubsection{\@startsection{subsubsection}{3}{\z@}%
                                   {-3.25ex\@plus -1ex \@minus -.2ex}%
                                   {1.5ex \@plus .2ex}%
                                   {\normalfont\normalsize\it}}
\renewcommand\paragraph{\@startsection{paragraph}{4}{\z@}%
                                   {-3.25ex\@plus -1ex \@minus -.2ex}%
                                   {1.5ex \@plus .2ex}%
                                   {\normalfont\normalsize\bf}}

\def\revise#1       {\raisebox{-0em}{\rule{3pt}{1em}}%
                     \marginpar{\raisebox{.5em}{\vrule width3pt\
                     \vrule width0pt height 0pt depth0.5em
                     \hbox to 0cm{\hspace{0cm}{%
                     \parbox[t]{4em}{\raggedright\footnotesize{#1}}}\hss}}}}

\def\del          {\partial}

\def\ee           {{\rm e}}

\def\tr           {\mathop{\rm Tr}}

\def\de#1#2{{\rm d}^{#1}\!#2\,}

\def\sqr#1#2{{\vcenter{\vbox{\hrule height.#2pt  
 \hbox{\vrule width.#2pt height#1pt \kern#1pt
 \vrule width.#2pt}\hrule height.#2pt}}}}

\def\a{\alpha}
\def\b{\beta}
\def\r{\rho}

\def\la{\lambda}

\def\be{\begin{equation}}
\def\ee{\end{equation}}

\def\m{\mu}
\def\g{\gamma}
\def\l{\lambda}
\def\n{\nu}

\catcode`\@=12

\begin{document}


   
\def \el {{\ell}}   
\def \KK {{\cal  K}}   
\def \K {{\rm K}}   
\def \tz{\tilde{z}}   
   
\def \ci {\cite}   
\newcommand{\rf}[1]{(\ref{#1})}   
\def \la {\label}   
\def \const {{\rm const}}

\def \ov {\over}   
\def \ha {\textstyle { 1\ov 2}}   
\def \we { \wedge}   
\def \P { \Phi} \def\ep {\epsilon}   
\def \ab {{A^2 \ov B^2}}   
\def \ba {{B^2 \ov A^2}}   
\def \tv   {{1 \ov 12}}   
\def \go { g_1}\def \gd { g_2}\def \gt { g_3}   
\def \gc { g_4}\def \gp { g_5}\def \F {{\cal F}}   
\def \del { \partial}   
\def \t {\theta}   
\def \p {\phi}   
\def \ep {\epsilon}   
\def \te {\tilde \epsilon}   
\def \ps {\psi}   
\def \x {{x_{11}}}

\def\br{\bar{\rho}}   
\newcounter{subequation}[equation]

\def\pa{\partial}   
\def\e{\epsilon}   
\def\rt{\rightarrow}   
\def\tr{{\tilde\rho}}   
\newcommand{\eel}[1]{\label{#1}\end{equation}}   
\newcommand{\bea}{\begin{eqnarray}}   
\newcommand{\eea}{\end{eqnarray}}   
\newcommand{\eeal}[1]{\label{#1}\end{eqnarray}}   
\newcommand{\LL}{e^{2\lambda(r)}}   
\newcommand{\NN}{e^{2\nu(r)}}   
\newcommand{\PP}{e^{-2\phi(r)}}   
\newcommand{\non}{\nonumber \\}   
\newcommand{\CR}{\non\cr}

\makeatletter   
   
\def\thesubequation{\theequation\@alph\c@subequation}   
\def\@subeqnnum{{\rm (\thesubequation)}}   
\def\slabel#1{\@bsphack\if@filesw {\let\thepage\relax   
   \xdef\@gtempa{\write\@auxout{\string   
      \newlabel{#1}{{\thesubequation}{\thepage}}}}}\@gtempa   
   \if@nobreak \ifvmode\nobreak\fi\fi\fi\@esphack}   
\def\subeqnarray{\stepcounter{equation}   
\let\@currentlabel=\theequation\global\c@subequation\@ne   
\global\@eqnswtrue \global\@eqcnt\z@\tabskip\@centering\let\\=\@subeqncr   
   
$$\halign to \displaywidth\bgroup\@eqnsel\hskip\@centering   
  $\displaystyle\tabskip\z@{##}$&\global\@eqcnt\@ne   
  \hskip 2\arraycolsep \hfil${##}$\hfil   
  &\global\@eqcnt\tw@ \hskip 2\arraycolsep   
  $\displaystyle\tabskip\z@{##}$\hfil   
   \tabskip\@centering&\llap{##}\tabskip\z@\cr}   
\def\endsubeqnarray{\@@subeqncr\egroup   
                     $$\global\@ignoretrue}   
\def\@subeqncr{{\ifnum0=`}\fi\@ifstar{\global\@eqpen\@M   
    \@ysubeqncr}{\global\@eqpen\interdisplaylinepenalty \@ysubeqncr}}   
\def\@ysubeqncr{\@ifnextchar [{\@xsubeqncr}{\@xsubeqncr[\z@]}}   
\def\@xsubeqncr[#1]{\ifnum0=`{\fi}\@@subeqncr   
   \noalign{\penalty\@eqpen\vskip\jot\vskip #1\relax}}   
\def\@@subeqncr{\let\@tempa\relax   
    \ifcase\@eqcnt \def\@tempa{& & &}\or \def\@tempa{& &}   
      \else \def\@tempa{&}\fi   
     \@tempa \if@eqnsw\@subeqnnum\refstepcounter{subequation}\fi   
     \global\@eqnswtrue\global\@eqcnt\z@\cr}   
\let\@ssubeqncr=\@subeqncr   
\@namedef{subeqnarray*}{\def\@subeqncr{\nonumber\@ssubeqncr}\subeqnarray}   
   
\@namedef{endsubeqnarray*}{\global\advance\c@equation\m@ne   
                           \nonumber\endsubeqnarray}   
   
\makeatletter \@addtoreset{equation}{section} \makeatother   
\renewcommand{\theequation}{\thesection.\arabic{equation}}   
   
\def \ci {\cite}   
\def \la {\label}   
\def \const {{\rm const}}   
\catcode`\@=11   
   
\newcount\hour   
\newcount\minute   
\newtoks\amorpm \hour=\time\divide\hour by 60\minute   
=\time{\multiply\hour by 60 \global\advance\minute by-\hour}   
\edef\standardtime{{\ifnum\hour<12 \global\amorpm={am}   
        \else\global\amorpm={pm}\advance\hour by-12 \fi   
        \ifnum\hour=0 \hour=12 \fi   
        \number\hour:\ifnum\minute<10   
        0\fi\number\minute\the\amorpm}}   
\edef\militarytime{\number\hour:\ifnum\minute<10 0\fi\number\minute}   
   
\def\draftlabel#1{{\@bsphack\if@filesw {\let\thepage\relax   
   \xdef\@gtempa{\write\@auxout{\string   
      \newlabel{#1}{{\@currentlabel}{\thepage}}}}}\@gtempa   
   \if@nobreak \ifvmode\nobreak\fi\fi\fi\@esphack}   
        \gdef\@eqnlabel{#1}}   
\def\@eqnlabel{}   
\def\@vacuum{}   
\def\marginnote#1{}   
\def\draftmarginnote#1{\marginpar{\raggedright\scriptsize\tt#1}}   
\overfullrule=0pt   
   
 \def \lc {light-cone\ }   
   
\def\draft{   
        \pagestyle{plain}   
        \overfullrule=2pt   
        \oddsidemargin -.5truein   
        \def\@oddhead{\sl \phantom{\today\quad\militarytime} \hfil   
        \smash{\Large\sl DRAFT} \hfil \today\quad\militarytime}   
        \let\@evenhead\@oddhead   
        \let\label=\draftlabel   
        \let\marginnote=\draftmarginnote   
        \def\ps@empty{\let\@mkboth\@gobbletwo   
        \def\@oddfoot{\hfil \smash{\Large\sl DRAFT} \hfil}   
        \let\@evenfoot\@oddhead}   
   
\def\@eqnnum{(\theequation)\rlap{\kern\marginparsep\tt\@eqnlabel}   
        \global\let\@eqnlabel\@vacuum}  }   
   
\renewcommand{\rf}[1]{(\ref{#1})}   
\renewcommand{\theequation}{\thesection.\arabic{equation}}   
\renewcommand{\thefootnote}{\fnsymbol{footnote}}   
   
\newcommand{\newsection}{    
\setcounter{equation}{0}   
\section}   
   
\textheight = 22truecm    
\textwidth = 17truecm    
\hoffset = -1.3truecm    
\voffset =-1truecm   
   
\def \tx {\textstyle}   
\def \tix{\tilde{x}}   
\def \bi{\bibitem}   
   
\def \ov {\over}   
\def \ha {\textstyle { 1\ov 2}}   
\def \we { \wedge}   
\def \P { \Phi} \def\ep {\epsilon}   
\def \ab {{A^2 \ov B^2}}   
\def \ba {{B^2 \ov A^2}}   
\def \tv   {{1 \ov 12}}   
\def \go { g_1}\def \gd { g_2}\def \gt { g_3}   
\def \gc { g_4}\def \gp {   
g_5}   
\def \F {{\cal F}}   
\def \del { \partial}   
\def \t {\theta}   
\def \p {\phi}   
\def \ep {\epsilon}   
\def \ps {\psi}

\def \LL{{\cal L}}   
\def\o{\omega}   
\def\O{\Omega}   
\def\e{\epsilon}   
\def\pd{\partial}   
\def\pdz{\partial_{\bar{z}}}   
\def\bz{\bar{z}}   
\def\e{\epsilon}   
\def\m{\mu}   
\def\n{\nu}   
\def\a{\alpha}   
\def\b{\beta}   
\def\g{\gamma}   
\def\G{\Gamma}   
\def\d{\delta}   
\def\r{\rho}   
\def\bx{\bar{x}}   
\def\by{\bar{y}}   
\def\bm{\bar{m}}   
\def\bn{\bar{n}}   
\def\s{\sigma}   
\def\na{\nabla}   
\def\D{\Delta}   
\def\l{\lambda}   
\def\te{\theta} \def \t {\theta}   
\def\ta {\tau}   
\def\na{\bigtriangledown}   
\def\p{\phi}   
\def\L{\Lambda}   
\def\hR{\hat R}   
\def\ch{{\cal H}}   
\def\ep{\epsilon}   
\def\bj{\bar{J}}   
\def \foot{ \footnote}   
\def\be{\begin{equation}}   
\def\ee{\end{equation}}   
\def \P {\Phi}   
\def\un{\underline{n}}   
\def\ur{\underline{r}}   
\def\um{\underline{m}}   
\def \ci {\cite}   
\def \g {\gamma}   
\def \G {\Gamma}   
\def \k {\kappa}   
\def \l {\lambda}   
\def \L {{L}}   
\def \Tr {{\rm Tr}}   
\def\apr{{A'}}   
\def \m {\mu}   
\def \n {\nu}   
\def \W{{\cal W}}   
\def \eps {\epsilon}   
\def \ha{{   
 { 1 \ov 2}} }   
\def \de{{   
{ 1 \ov 9}} }   
\def \si{{   
 { 1 \ov 6}} }   
\def \fo{{   
{ 1 \ov 4}} }   
\def \ei{{   
{ 1 \ov 8}} }   
\def \rt {{\tx { \ta \ov 2}}}   
\def \rr {{\bar \rho}}   
   
\def\D{\Delta}   
\def\l{\lambda}   
\def\L{\Lambda}   
\def\te{\theta}   
\def\g{\gamma}   
\def\Te{\Theta}   
\def\tw{\tilde{w}}

\def\sn{\rm sn} 
\def\cn{\rm cn}  
\def\dn{\rm dn}

\def\hzero{\hat{0}}
\def\ha{\hat{a}}
\def\hb{\hat{b}}
\def\hc{\hat{c}}
\def\hd{\hat{d}}
\def\he{\hat{e}}

\def\hone{\hat{1}}
\def\htwo{\hat{2}}
\def\hthree{\hat{3}}
\def\hz{\hat{z}}
\def\hteone{\hat{\theta}_1}
\def\htetwo{\hat{\theta}_2}
\def\hpone{\hat{\phi}_1}
\def\hptwo{\hat{\phi}_2}
\def\hpsi{\hat{\psi}}


\topmargin=0.50in  
   
\date{}   
   
\begin{titlepage}   
   
\version\versionno  
   
\hfill hep-th/0410035    
   
\hfill MCTP-04-56

\hfill BRX TH-552
   
\begin{center}   

{\Large \bf Regge Trajectories for Mesons}\\

\vskip .3cm 
{\Large \bf  in the Holographic Dual of  Large-$N_c$
QCD }

\vskip .7cm

Mart\'\i  n Kruczenski$^{1}$,  Leopoldo A. Pando Zayas$^{2}$\\
Jacob Sonnenschein$^3$  and Diana Vaman$^{2,4}$   
   

\end{center}   

\vskip .4cm \centerline{\it ${}^1$ Department of Physics}   
\centerline{ \it Brandeis University,  Waltham, MA 02454}    
   
\vskip .2cm \centerline{\it ${}^2$ Michigan Center for Theoretical   
Physics}   
\centerline{ \it The University of   
Michigan, Ann Arbor, MI 48109-1120}    

\vskip .2cm 
\centerline{\it ${}^3$ School of Physics and Astronomy}    
\centerline{ \it Beverly and Raymond Sackler Faculty of Exact Sciences}    
\centerline{ \it Tel Aviv University, Ramat Aviv, 69978, Israel}

\vskip .2cm \centerline{\it ${}^4$ Department of Physics }   
\centerline{ \it Princeton University,  Princeton, NJ 08544}

\vskip .5 cm

\begin{abstract}   
We discuss Regge trajectories of dynamical mesons in large-$N_c$ QCD, 
using the supergravity background describing $N_c$ D4-branes
compactified on a thermal circle. The
flavor degrees of freedom arise from the addition of
$N_f<<N_c$ D6 probe branes. Our work provides a string theoretical 
derivation, via the gauge/string correspondence, 
of a phenomenological model describing the meson  
as rotating point-like massive particles connected by a flux 
string. The massive endpoints induce nonlinearities for the Regge
trajectory. For light quarks the Regge trajectories of mesons are
essentially linear. For massive quarks our trajectories qualitatively capture the
nonlinearity detected in lattice calculations. 
\end{abstract}

\end{titlepage}   
\setcounter{page}{1} \renewcommand{\thefootnote}{\arabic{footnote}}   
\setcounter{footnote}{0}   
   
\def \N{{\cal N}}    
\def \ov {\over}

\section{Introduction and Summary}  

The Regge trajectories of mesons were among the first indications that
mesons admit a  stringy behavior. Indeed, a very elementary treatment
of a classical spinning open bosonic  string in flat space time yields
a simple  relation between the angular momentum and energy, $J=
\alpha' E^2$, which is a Regge trajectory.  Quantum corrections of
the classical bosonic string  configuration  add an intercept to the
trajectory, namely  $J= \alpha' E^2 +  \frac{(d-2) \pi}{12}+
O(E^{-2})$.

In the modern era of the gauge/gravity holographic duality one wonders
whether a similar behavior characterizes  the classical spinning
string in  supergravity backgrounds that  are  duals of  confining
gauge theories.   In \cite{PSV} folded closed spinning strings
were analyzed in the context of the KS and MN backgrounds. It was
found that classically these  configurations indeed admit a linear
Regge behavior $J = \alpha'_{eff} E^2$  where $\alpha'_{eff}=
\frac{\alpha'}{g_{00}(\hat U)}$  and where  $\hat U$ denotes the
location of the spinning string along the radial direction.
Interestingly, it was found that the  quantum corrected trajectory is
different  than that of the bosonic string in flat space time, it
takes the form  $J= \alpha'_{eff} E^2 + \alpha_0+ \beta E $ where
$\alpha_0$ is not a pure number but rather depends on the parameters
of the background, and so is $\beta$.  These configurations map into
the  Regge trajectories of glueballs in the dual gauge  theory.
Glueball Regge trajectories with nonvanishing intercept were also
found in lattice  YM  calculations \cite{Teper}.

There are several models of supergravity backgrounds holographically
dual to confining gauge theories.  In this paper  we will make use of
the one 
associated with  the near extremal $N_c$ D4 branes \cite {wqcd}. By
imposing anti-periodic boundary conditions along the  thermal circle for the fermions,
the background corresponds to  the low energy regime of the pure YM
theory in four dimensions. The stringy Wilson loop in this background
was computed in \cite{BISY} using the methods of \cite{wl}. Whereas
the Regge behavior of glueballs
can be addressed in this model \cite{Leo}, mesonic spinning strings
cannot be described in this framework since its dual theory  does not
include quarks in the fundamental representations.  Flavor fundamental
quarks and anti-quarks can be invoked  in  supergravity backgrounds by
introducing flavor probe branes \cite{karch}.  Probe  D7 branes were
incorporated in models based on $D3 $ branes in
\cite{spinning,SS,flavorwqcd,bjlv}. Similar consideration of flavor can
be found in \cite{flavor}. 
The idea behind  some these papers has been to  extract the
mesonic spectrum from the fluctuations of the  fields, like pseudo
scalars and vectors, on the flavor probe branes.  This, obviously,
cannot be done in supergravity for states with spin higher than  two. To describe the
Regge trajectories associated with high spins, one is naturally
forced to redo the ``old'' calculations of spinning strings, but now
not in the context of the bosonic string in 4d but rather in the
context of the supergravity confining backgrounds that include probe branes.
Concretely we followed in this work the proposal of \cite{flavorwqcd} of
adding D6 probe branes to the background of the  near extremal $N_c$
D4 branes.

By  analyzing the conditions of having non trivial spinning strings,
we show that the conditions of having spinning strings at a  constant
radial coordinate, resembles the conditions of having an area law
Wilson line \cite{Kinar} and that it can occur only along the `` wall
of space'' in the radial direction.

An interesting result of our analysis is the fact that the solution of
the spinning strings is very closely related to  a toy model of a
string with two masses attached to its endpoints, which spins in flat
space-time. This simple model was extensively considered in the 70's
by, for example, \cite{oldmass,stringtoy} and it has been reviewed in
the book \cite{stringbook} where further references can be found. More
importantly, this model which is also referred to  as the mass-loaded
generalization of the Chew-Frautschi formula, is an essential tool in
the extremely successful
approach to meson and hadron spectroscopy developed recently by Wilczek
\cite{wilczek}. 

We describe the toy model and solve it prior to performing
the stringy calculations. The main outcome  of this paper is that the classical spinning strings
with endpoints on the probe branes, admit   Regge trajectories which
get corrections due the ``masses of the quarks''
\begin{eqnarray}
E=\frac{2T_g}{\omega}\left(\arcsin x +\frac{1}{x}\sqrt{1-x^2}\right),
\qquad
J= \frac{T_g}{\omega^2}\left(\arcsin  x + \frac{3}2 x\sqrt{1-x^2}\right),
\end{eqnarray}
where $x$ is the speed of the endpoints of the strings and the mass of
the quarks is $m_q=T_g (1-x^2)/(\omega x)$. The Regge
regime requires relativistic motion of the string, that is, $x\to1$. 
The massive endpoints induce nonlinearities for the Regge
trajectory. For light quarks the Regge trajectories of mesons are
essentially linear. For massive quarks our trajectories qualitatively capture the
nonlinearity detected in lattice calculations. 
In particular, the  slope for the lowest states within the Regge
trajectories of heavy-quark mesons is flavor dependent, while for
the highest states the slope is universal. We explain the universality
of the second slope in terms of  generic properties of confining
holographic backgrounds.

The organization of the paper is as follows. In section 2 we describe
the general setup of spinning strings in confining
backgrounds. Section 3 is devoted to reviewing a toy model consisting
of two massive relativistic particles attached to the ends of  a
spinning string. The energy  and angular momentum of the system are
computed and are shown to admit a Regge trajectory behavior with
corrections that depend on mass of the particles. In section 4 we
analyze the Regge trajectories of mesons from the perspective of the 
gauge/string duality.  We investigate a macroscopic spinning open
string configuration, whose endpoints are located at the boundary of
the flavor brane. It is shown that one can approximate its
configuration with that of two vertical strings stretched between the
flavor brane and the ``floor'' and a horizontal string that stretches
along the wall.  This classical configuration has a Regge like
trajectory which indeed   can be mapped into that  of the toy model.
Finally, in section 5 we comment on the rapport between the toy model
and the more recent literature on the phenomenology of mesons. 

\section{ Spinning open strings in confining backgrounds: General  setup} 
Consider a background metric of the form \be  ds^2 = -g_{00}(U)  dt^2
+ g_{ii}(U) ({dx^{i}})^2 + g_{UU} dU^2 + ...  \ee where $x^{i}$ are
the space coordinates associated with the  uncompactified worldvolume,
$U$ is a radial coordinates and ellipsis  stands for additional
transverse directions. The metric as well as any other field of the
background depend only on the radial direction $U$.

In \cite{Kinar} it was shown that backgrounds with metric of this form admit
an area law behavior, namely confinement, if one of the two conditions is obeyed
\bea\label{conf}
&& g_{00}g_{ii}(U){\rm \ has\ a \ minimum\  at\  }\, U=U_{min}  \qquad {\rm and} \qquad g_{00} g_{ii}(U_{min})> 0, \\ 
&&g_{00} g_{UU} |_{ U= U_\Lambda} \rightarrow  \infty  \qquad {\rm and} \qquad g_{00} g_{ii}(U_\Lambda)> 0. 
\eea

Backgrounds that obey the first condition  are for instance  the
KS and  the MN  backgrounds and their
non-supersymmetric  deformations. For the metric of the
$AdS_5\times S^5$ black hole and its non extremal $D_p$ analogs the
second condition is obeyed.   In both cases the  string that
associates with the Wilson line stretches vertically from the
boundary  to the ``wall''( or  ``end of space''  or   ``floor'')
where it stretches horizontally and then up vertically to the
boundary. In the horizontal segment the string is in fact that of  a
flat space time and hence the area law behavior. We will see belove
that the condition of having a spinning string with a constant radial
coordinate will be similar to those of having confining Wilson loop,
and the string will spin along the ``wall''.

Since we have in mind addressing spinning strings, it is convenient to describe
the space part of the metric as 
\be
dx_{i}^2 = dR^2 + R^2 d\t ^2 + dx_3 ^2, 
\ee
where $x_3$ is the direction perpendicular to the plan of rotation. 
The classical equations of motion of a bosonic string defined on this background can be
formulated on  equal footing in the NG formulation or the  Polyakov action.
Let us use now the latter.  The equations  of motion 
associated with the variation of $t, \t, R $ and $U$  respectively are  
\bea
&&\pa_\alpha ( g_{00}\pa^\a t )  = 0 \\ \nonumber
&&\pa_\alpha ( g_{ii} R^2\pa^\a\t )  = 0 \\ \nonumber
&&\pa_\alpha ( g_{ii}\pa^\a R) - g_{ii} R \pa_\a \t \pa^\a \t  = 0  \\ \nonumber
&&2\pa_\alpha ( g_{UU}\pa^\a U) + \frac{d g_{00}}{dU}  \pa_\a t \pa^\a t 
-\frac{d g_{ii}}{dU}  \pa_\a x^i \pa^\a x^i -  \frac{d g_{UU}}{dU}  \pa_\a u \pa^\a u 
= 0, \\
\eea
where $\a$ denotes the worldsheet coordinates $\tau$ and $\sigma$.
In addition in the Polyakov formulation one has to add the Virasoro constraint
\be
-g_{00} (\pa_{\pm} t)^2+g_{ii} (\pa_{\pm} x^i)^2 + g_{UU} (\pa_{\pm} U)^2 + ... =0,
\ee
where $\pa_{\pm}= \pa_{\tau}\pm \pa _ \sigma$ and  $...$ stands for the contribution
to the Virasoro constraint of the rest of the background metric. 

Next we would like to find solutions of the equations of motion which describe strings   
spinning in space-time. For that purpose we take the following ansatz
\be\label{spinning}
t = \tau \qquad  \t = \omega \tau \qquad R(\sigma\tau) = R(\sigma)\qquad U=\hat U= {\rm constant}.
\ee
It is obvious that this ansatz solves the first two equations. The third equation
together with the Virasoro constraint is  solved ( for the case that $g_{00}=g_{ii}$) by  
$R= A cos(\omega\sigma) + B sin(\omega\sigma)$  with $\omega^2 ( A^2 + B^2) =1$.  The
boundary conditions 
we want to impose will select 
the particular combination of $A$ and $B$. 
Let us now investigate the equation of motion associated with $U$ and for the particular ansatz  $U=\hat U$.
This can be  a solution only provided
\be\label{condition}
\frac{d g_{00}}{dU}|_{U= \hat U}= 0 \qquad  \frac{d g_{ii}}{dU}|_{U= \hat U}=  0. 
 \ee
This is just the first condition for having a confining background. The condition 
$ g_{00} g_{ii}(\hat U)> 0$ insures that the Virasoro constraint is obeyed in a non trivial manner. 
In \cite{PSV} spinning strings in the KS and MN models, which belong to
this class of confining 
backgrounds, were analyzed.

In the background that we will employ in the  present paper, the near
extremal $D4$ brane, one 
direction is along an $S^1$ 
and  there is   no five dimensional Lorentz invariance. The direction
along the $S^1$ denoted 
by $\psi$ has a metric 
which is  related to that of the $U$ direction as follows 
$g_{\psi\psi }= [g_{UU}]^{-1}$. Thus the  condition of having a
solution of the equation of 
motion with $U=\hat U$ 
includes the condition 
\be
\frac{d g_{\psi\psi}}{dU}|_{U= \hat U}= -\frac{\frac{d g_{UU}}{dU}}{g_{UU}^2}|_{U= \hat U}= 0. 
\ee
 This condition is obeyed if at $U=\hat U$  $g_{UU}(\hat U)\rightarrow \infty$ which 
 is the second condition for a confining background (\ref{conf}) with $\hat U= U_\Lambda$, and again with the 
the demand of non-vanishing $ g_{00} g_{ii}(\hat U)> 0$ to have a non-trivial Virasoro constraint. 
The left over part of the equation of motion (\ref{condition}) will be shown to be obeyed  in section 4 by 
 transforming the radial coordinate  to a coordinate  that measures the distance from $U_\Lambda$.

To summarize, we have just realized that there is a close relation between the conditions of having area law 
Wilson loop and of having a spinning string configuration at a constant radial coordinate. 
\section{A toy model of the meson}\label{toy}
Prior to analyzing the string configuration that describes a meson, let first 
review a toy model that consists of of two massive, relativistic particles
connected by a string (see figure \ref{toymodel}):

\begin{figure}
\begin{center}
\epsfig{file=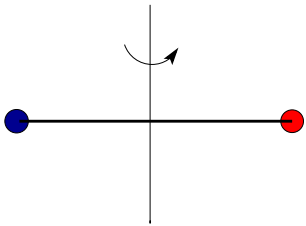,width=4cm}
\caption{The toy model. \label{toymodel}}
\end{center}
\end{figure}

Such models were proposed in the past as an effective description
of mesons \cite{stringtoy,stringbook, bura}. However, the reason we
address them here is that they are intimately related
to the stringy meson  associated with a confining background, as we will 
show in the next section. 

Consider the flat space-time 
 NG action of an open string combined with the action of  two relativistic particles of equal mass $m$ attached to 
the endpoints of the string.

\bea\label{toyaction}
S &=& -T \int d\tau \int_{-\frac{\pi}{2}}^{\frac{\pi}{2}}d\s \sqrt{ -[ (\dot{t})^2- \dot R^2 - R^2 \dot\t^2]
[ ( t')^2-  {R'}^2 - R^2 {\t '}^2]+ [-\dot{t}t'+\dot{R}R'+R^2\dot{\theta}\theta']^2}\nonumber\\
\\ 
& -&m \int d\tau \sqrt{ -[ (\dot t)^2- \dot R^2 - R^2 \dot\t^2]}\big|_{\sigma=-\frac{\pi}{2}} -m\int d\tau 
\sqrt{ -[ (\dot t )^2- \dot R^2 - R^2 \dot\t^2]}\big|_{\sigma=\frac{\pi}{2}},\nonumber\\
\eea 
where the metric was taken to be
$ds^2 = -dt^2 + dR^2 + R^2 d\t^2$ is the space-time metric relevant to the rotating string,
 $T$ is the effective string tension 
$m$ is the mass of the particles, the world volume coordinate $\sigma$ is defined in the interval 
$[-\frac{\pi}{2},\frac{\pi}{2}]$
and  $\dot A$ and $A'$  denote derivatives  with respect to $\tau$ and $\sigma$ 
respectively.  

Note that in the string part of the action $X^0, R\  {\rm and}\  \t$
are {\it a priori} functions
of $\sigma$ and $\tau$ and in the particle only of $\tau$. Moreover, due to the reflection symmetry of the problem, we take that the 
values of $t, R\  {\rm and}\  \t$ at $\sigma$ equal those of $t, R\  {\rm and}\  \t$ at $-\sigma$.
Next we discuss the equations of motion. Due to the fact that there are particle actions at 
the end points of the interval of $\sigma$, the equations of motion 
of $x^\mu$ will include a bulk equation and a surface one as can be see from the following variation
\bea
\delta S &=& -T \int d \tau  \int_{-\frac{\pi}{2}}^{\frac{\pi}{2}}d\s
\left [\frac{\pa{\cal L}_s}{\pa X^\mu  }- \pa_\a [ \frac{\pa{\cal
	L}_s}{\pa_\a {X^\mu}  }] \right ]
\delta X^\mu(\s,\tau)\nonumber
\\
&+& \int d \tau\left [ T   \frac{\pa{\cal L}_s}{ \pa_\s {X^\mu}  }
\delta X^\mu(\s,\tau) \big|_{\sigma={-\frac{\pi}{2}}} 
- m [\pa_\tau[ \frac{\pa{\cal L}_p}{\pa_\tau {X^\mu}  }] -  
\frac{\pa{\cal L}_p}{\pa X^\mu }]\delta X^\mu(\tau)\right ] \nonumber 
\\
&-& \int d \tau\left [ T   \frac{\pa {\cal L}_s}{ \pa_\s {X^\mu}
  }\delta X^\mu(\s,\tau)\big|_{\sigma
={\frac{\pi}{2}}} 
- m [\pa_\tau[ \frac{\pa{\cal L}_p}{ \pa_\tau {X^\mu}  }]
- \frac{\pa{\cal L}_p}{\pa X^\mu }]\delta X^\mu(\tau) \right ].\nonumber
\\
\eea
Thus the equations of motion are 
\bea 
\frac{\pa {\cal L}_s}{\pa X^\mu  }- \pa_\a[ \frac{\pa{\cal L}_s}{ \pa_\a {X^\mu}  }] &=& 0, \nonumber
\\
T   \frac{\pa{\cal L}_s}{ \pa_\s {X^\mu}
}\big|_{\sigma={-\frac{\pi}{2}}} 
- m \pa_\tau[ \frac{\pa{\cal L}_p}{ \pa_\tau {X^\mu}  }] 
+ m \frac{\pa{\cal L}_p}{ \pa {X^\mu}  }&=&0,\nonumber
\\
T   \frac{\pa {\cal L}_s}{ \pa_\s {X^\mu}  }\big|_{\sigma={\frac{\pi}{2}}} - m \pa_\tau[ \frac{\pa{\cal L}_p}{ \pa_\tau {X^\mu}  }]+ m \frac{\pa{\cal L}_p}{ \pa {X^\mu}  }&=&0.\nonumber
\\
\eea
We now look for spinning configurations based on the following ansatz (\ref{spinning}) 
\be
t= \tau, \qquad \t = \omega \tau, \qquad R(\tau,\s) = R( \s).
\ee
For this type of ansatz the bulk equations of motion are
\bea
\pa _\tau \left ( \frac {  R'(\s)}{\sqrt{ 1- \omega^2 R^2(\s)}}\right ) &=& 0, \nonumber
\\
\pa _\tau \left ( \frac { \omega R'(\s)}{\sqrt{ 1- \omega^2 R^2(\s)}}\right ) &=& 0, \nonumber
\\ 
\pa _\s \left ( \sqrt{ 1- \omega^2 R^2(\s)}\right ) + \frac{\omega^2 R R'(\s) }{\sqrt{ 1- \omega^2 R^2(\s)} } &=& 0. 
\eea
The first and second equations are trivially obeyed. The third equation is obeyed in fact for any $R(\s)$.
The particular form of it depends on the boundary conditions that one wants to impose. For instance, for the closed string,
  requiring the 
periodicity $R(\s)= R(\s+ 2\pi)$,  where the range of $\s$ is taken to be $(0,2\pi)$, 
 is compatible with a solution of the form $R= \frac{1}{\omega} \cos (\s)$.
Or, for the case of rotation Wilson line one imposes $R(0)=0$ and $R(\pi)=L$ and the corresponding solution 
$R= L sin (\s/4)$.

In the present case the boundary conditions will be determined by the surface equations of motions. In fact the surface equations for $t $
and $\t$ are trivial, and only the equations for $R$ constrain the solutions. The latter reads
\be
T \left ( \frac{\pa {\cal L}_s}{\pa R'}\right )\bigg|_{R=R_0}   + m\left (   \frac{\pa {\cal L}_p}{\pa R}\right ) = 0,  \nonumber
\ee
which implies
\be\label{sew}
T \sqrt{ 1- \omega^2 R_0^2} = m   \frac{R_0\omega^2 }{\sqrt{ 1- \omega^2 R_0^2}},
\ee
where  $R_0\equiv R(-\pi/2)=R(\pi/2)$.
This expression enables us to determine $\omega$ in terms of $R_0$ or vice-versa. 
For the case of $m=0$, namely without the action of the particles,  the surface term of the string action should vanish. 
Note that the same relation can be derived  by minimizing the action with respect to $R_0$.

Next we want to find the energy and the angular momentum of the system.
 Starting with the action \ref{toyaction}  the energy and angular momentum are given by
\bea
E &=& T \int_{\frac{\pi}{2}}^{\frac{\pi}{2}}d\s \frac{\dot t \sqrt{[ ( t')^2-  {R'}^2 - R^2 {\t '}^2]}}{ \sqrt{ -[ (\dot t)^2- \dot R^2 - R^2 \dot\t^2]}}
+m \frac{\dot t }{ \sqrt{ -[ (\dot t)^2- \dot R^2 - R^2 \dot\t^2]}}\big|_{\s=-\pi/2,\pi/2}\nonumber
\\
&=&2  T\int_0^{R_0} dR \frac{1}{\sqrt{ 1- \omega^2 R^2(\s)}} +2 m
 \frac{1}{\sqrt{ 1- \omega^2 R_0^2}} =  
2 \frac{T}{\omega}\arcsin( \omega R_0)+2 m \frac{1}{\sqrt{ 1- \omega^2 R_0^2}}\nonumber
\eea
\bea
J &=& T \int_{\frac{\pi}{2}}^{\frac{\pi}{2}}d\s \frac{R^2\dot \t \sqrt{[ ( {t}')^2-  {R'}^2 - R^2 {\t '}^2]}}{ \sqrt{ -[ (\dot t )^2- \dot R^2 - R^2 \dot\t^2]}}
+m \frac{R_0^2\dot \t }{ \sqrt{ -[ (\dot t)^2- \dot R^2 - R^2 \dot\t^2]}}\big|_{\s=-\pi/2,\pi/2}\nonumber
\\
&=& 2 T\omega\int_0^{R_0} dR \frac{R^2}{\sqrt{ 1- \omega^2 R^2(\s)}} +2 m \frac{R_0^2 \omega}{\sqrt{ 1- \omega^2 R_0^2}} \nonumber
\\
& = &  \frac{T}{\omega^2}
\left [\arcsin( \omega R_0)- \omega R_0\sqrt{ 1- \omega^2 R^2(\s)}\right ]+ 2 m \frac{\omega R_0^2}{\sqrt{ 1- \omega^2 R_0^2}}.\nonumber
\eea
 So  we can summarize the results as follows
\bea
E &=& 2 \frac{T}{\omega}[ \arcsin( \omega R_0) + 2 \sqrt{\frac{m}{TR}}],\nonumber
\\
J &=& \frac{T}{ \omega^2}[ \arcsin( \omega R_0) + 3(\omega R_0)^2 \sqrt{\frac{m }{TR_0}}].
\eea

Obviously for $m=0$ we are back to the standard linear Regge trajectory. 
It is straightforward to generalize the case of a string with  equal
masses
at his
endpoints to the case where at each endpoint there are two different
masses.
In this case we get two "sewing" conditions instead of (\ref{sew}) and
there
will be contributions to the energy and J from the two masses expressed in
terms of two separation distances.

In the next section the relation between this toy model and the stringy meson will be made clear
 by deriving a similar relation between $J$ and $E$.

\section{The meson as a rotating string}

In this section we derive the Regge trajectory of a meson with large
spin, from the perspective of the gauge/string duality.
For specificity the background we consider is the supergravity background
of a large number $N_c$ of D4 branes compactified on a thermal circle. By adding 
$N_f<<N_c$ D6 probe branes 
(by probe we mean that we ignore the backreaction of the D6
branes),  this geometry
becomes the holographic dual of 4 dimensional large-$N_c$ 
QCD, with $N_f$ flavors plus KK modes.   
Without loss of generality we restrict in the following to the case when 
$N_f=1$.
A meson with large spin will correspond to 
an open spinning string string in the supergravity background sourced by the
the non-extremal D4 branes in the decoupling limit
\be
ds^2=\frac{U^{3/2}}{R_{D4}^{}{3/2}}(dX^\mu dX^\nu \eta_{\mu\nu}+ 
f(U)d\psi^2)+K(U)(d\rho^2+\rho^2 d\Omega_4^2),
\ee
where
\bea
f(U)&=&1-\frac{U_{\Lambda}{}^3}{U^3},\nonumber \\
U(\rho)&=&(\rho^{3/2}+\frac{U_{\Lambda}^3}{4\rho^{3/2}})^{2/3},
\nonumber\\
K(U)&=&R_{D4}{}^{3/2}U^{1/2}\rho^{-2}.\label{uu}
\eea
The coordinate $\psi$ parametrizes the thermal circle on which the 
D4 branes are compactified, and $X^\mu$ with $ \mu=0,1,2,3$ 
are coordinates in the remaining
4 non-compact directions along the D4 branes. 
The endpoints of this macroscopic string are located on the D6 probe brane
which extends in the $\rho$ direction from the boundary $\rho=\infty$ to
$\rho=\rho_{f}$.

\begin{figure}
\begin{center}
\epsfig{file=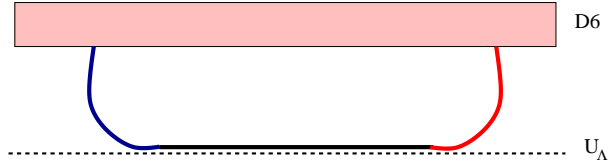,width=8cm}
\caption{String ending on the probe D6. \label{string}}
\end{center}
\end{figure}

More precisely, the D6 probe spans the same 4 non-compact directions
$X^\mu$ as the D4 branes, plus 3 other non-compact 
directions in the transverse space
to the D4, $\lambda^i , i=1,2,3$: 
\be
ds_5{}^2= d\rho^2+\rho^2 d\Omega_4{}^2=d\lambda^i d\lambda^i+dr^2+r^2 
d\phi^2=d\lambda^2 +\lambda^2 d\Omega_2^2+dr^2+r^2 
d\phi^2. 
\ee
Solving the equations of motion derived from the Born-Infeld action
\cite{flavorwqcd}, the D6 probe profile $r=r(\lambda), \phi=$constant 
is described by the equation
\be 
\frac{d}{d\lambda}\bigg[ \bigg( 1+\frac{U_{\Lambda}{}^3}{4\rho^3}\bigg)^2
\lambda^2 \frac{dr/d\lambda}{\sqrt{1+(dr/d\lambda)^2}}\bigg]=-\frac{3}{2}
\frac{U_{\Lambda}{}^3}{\rho^5}\bigg(1+\frac{U_{\Lambda}{}^3}{4\rho^3}\bigg)
\lambda^2 \frac{dr/d\lambda}{\sqrt{1+(dr/d\lambda)^2}}.
\ee
Due to the non-trivial profile of the D6 brane, the $U_A(1)$ symmetry is
spontaneously broken and the quark condensate $\langle \bar q q\rangle$
acquires a vev. Asymptotically this symmetry is restored, 
$r=\rm{constant}+\frac{c}{\lambda}$ and $c$ is related to the quark condensate
\cite{flavorwqcd}. 

In terms of the extent of the D6 brane probe in the 5 dimensional space
transverse to the D4 branes, we have
$
\rho_{D6}^2=r(\lambda)^2+\lambda^2
$. Thus, as advertised, the D6 stretches in the $\rho$ directions
from $\rho_{f}$ at $\lambda=0$\footnote{It is easy to see that
$d\rho/d\lambda=2(\lambda + r (dr/d\lambda))=0$ at $\lambda=0$.}
to $\infty$ at $\lambda\to \infty$.

We now return to the description of a meson of dynamical quarks as a 
rotating open string whose endpoints lie on the D6 probe. 
We begin with the  ansatz
\be
X^0=e\tau, \quad \theta= e\omega\tau, \quad R=R(\sigma), \quad r=r(\sigma),
\quad \lambda=\lambda(\sigma), \label{ansatz}
\ee 
where we parametrized the 4d non-compact space on the worldvolume
of the non-extremal D4 branes by
\be
dX^\mu dX_\mu=-(dX^0){}^2+ dR^2+ R^2 d\theta^2 + (dX^3){}^2.
\ee
The boundary conditions  corresponding to this configuration
are Neumann in the directions parallel with the D6 brane, and Dirichlet
in the directions transverse to the brane:
\bea
&&\bigg(
\partial_\sigma X^0=\partial_\sigma\theta=\partial_\sigma R=\partial_\sigma 
\lambda\bigg)\bigg|_{\sigma=0, \pi}=0,\nonumber\\
&& r(\sigma)=r(\lambda(\sigma))\bigg|_{\sigma=-\pi/2,\pi/2}.
\eea
Notice that the boundary conditions are trivially satisfied in the
$X^0$ and $\theta$ coordinates. 
Plugging this ansatz in the NG action we search for a solution 
characterized by $\lambda={\rm constant}$. The equation of motion associated with $\lambda$ 
\bea
&&\partial_\sigma\bigg(\frac{K \sqrt{(U/R_{D4})^{3/2}
(\partial_\tau X^0{}^2 - R^2 
\partial_\tau \theta{}^2)}}
{\sqrt{(U/R_{D4})^{3/2}\partial_\sigma R{}^2+K(\partial_\sigma
r{}^2 + \partial_\sigma \lambda{}^2)}} \partial_\sigma\lambda \bigg)\nonumber\\
&=&\frac{\lambda}{\rho}
\frac{d}{d\rho}\bigg(\sqrt{(U/R_{D4})^{3/2}(\partial_\tau X^0{}^2 - R^2 
\partial_\tau \theta{}^2)((U/R_{D4})^{3/2}
\partial_\sigma R{}^2+K(\partial_\sigma
r{}^2 + \partial_\sigma \lambda{}^2)} \bigg), \nonumber
\eea
is trivially satisfied for $\lambda=0$.
The picture that transpires out of this analysis is that of a spinning
 open string whose endpoints are located at the termination point of the 
D6 brane in the $\rho$ direction. 

Adding $\lambda=0$ to the ansatz (\ref{ansatz}), the NG action reads:
\be
S=- T_s\int d\sigma d\tau \sqrt{(U/R_{D4})^{3/2}
(\partial_\tau X^0{}^{2}- R^2 \partial_\tau
\theta{}^2)((U/R_{D4})^{3/2}\partial_\sigma R{}^2 + 
K\partial_\sigma \rho{}^2)}, 
\label{act}
\ee
where we used that for $\lambda=0$ we have $r=\rho_{D6}$. From the variational principle
\bea
0&=&\delta S\nonumber\\
&=&
\int d\sigma d\tau \;\delta R\bigg[-\partial_\sigma\bigg(\frac{
(U/R_{D4})^{3/2}
\partial_\sigma R \sqrt{(\partial_\tau X^0{}^{2}- R^2 \partial_\tau
\theta{}^2)(U/R_{D4})^{3/2}}}{\sqrt{(U/R_{D4})^{3/2}\partial_\sigma R{}^2 + 
K\partial_\sigma \rho{}^2}}\bigg)
\nonumber\\
&-&\frac{R\partial_\tau\theta{}^2\sqrt{
(U/R_{D4})^{3/2}((U/R_{D4})^{3/2}\partial_\sigma R{}^2 + 
K\partial_\sigma \rho{}^2)}}{
\sqrt{\partial_\tau X^0{}^{2}+ R^2 \partial_\tau
\theta{}^2}}\bigg]\nonumber\\
&+&\int d\sigma d\tau \; \delta \rho
\bigg[-\partial_\sigma\bigg(\frac{K\partial_\sigma\rho
\sqrt{(\partial_\tau X^0{}^{2}- R^2 \partial_\tau\theta{}^2)(U/R_{D4})^{3/2}}}
{\sqrt{(U/R_{D4})^{3/2}\partial_\sigma R{}^2 + K\partial_\sigma \rho{}^2}}
\bigg)
\nonumber\\
&+&\frac{\sqrt{\partial_\tau X^0{}^{2}- R^2 \partial_\tau
\theta{}^2}(3\frac{dU}{d\rho}R_{D4}^{-1}
((U/R_{D4})^2 \partial_\sigma R{}^2+1/2 (U/R_{D4})^{1/2} 
K \partial_\sigma\rho{}^2)+ (U/R_{D4})^{3/2} 
\frac{dK}{d\rho}\partial_\sigma\rho{}^2)}{2\sqrt{
(U/R_{D4})^{3/2}((U/R_{D4})^{3/2}\partial_\sigma R{}^2 
+ K\partial_\sigma \rho{}^2)}}\bigg]\nonumber\\
&-&\int d\sigma d\tau \;\delta X^0\partial_\tau\bigg(\frac{\partial_\tau X^0 
\sqrt{(U/R_{D4})^{3/2}((U/R_{D4})^{3/2}\partial_\sigma R{}^2 
+ K\partial_\sigma \rho{}^2)}}{\sqrt{\partial_\tau X^0{}^{2}- R^2 
\partial_\tau\theta{}^2}}\bigg)\nonumber\\
&+&\int d\sigma d\tau \;
\delta\theta\partial_\tau\bigg(\frac{R^2\partial_\tau\theta
\sqrt{(/R_{D4})U^{3/2}((U/R_{D4})^{3/2}\partial_\sigma R{}^2 + 
K\partial_\sigma \rho{}^2)}}{\sqrt{\partial_\tau X^0{}^{2}- R^2 \partial_\tau
\theta{}^2}}\bigg)\nonumber\\
&+&\int d\tau \delta R\frac{(U/R_{D4})^{3/2}
\partial_\sigma R \sqrt{(\partial_\tau X^0{}^{2}- R^2 \partial_\tau
\theta{}^2)(U/R_{D4})^{3/2}}}{\sqrt{(U/R_{D4})^{3/2}\partial_\sigma R{}^2 + 
K\partial_\sigma \rho{}^2}}\bigg|_{\sigma=-\pi/2}^{\sigma=\pi/2}
\nonumber\\
&+&\int d\tau \delta \rho\frac{K\partial_\sigma\rho
\sqrt{(\partial_\tau X^0{}^{2}- R^2 \partial_\tau\theta{}^2)(U/R_{D4})^{3/2}}}
{\sqrt{(U/R_{D4})^{3/2}\partial_\sigma R{}^2 + 
K\partial_\sigma \rho{}^2}}\bigg|_{\sigma=-\pi/2}^{\sigma=\pi/2}.
\nonumber\\\label{deltas}
\eea
we extract the equations of motion and boundary conditions.
As we have already mentioned, we impose Dirichlet boundary conditions
in $\rho$ and Neumann boundary conditions in $R$. Combining these two 
conditions into a single one we find that the string ends transversely
to the probe brane, i.e.
$d\rho/dR=d\rho/d\sigma \cdot d\sigma/dR=\infty$ 
at the endpoints of the string.

The equation of motion derived from the NG action is 
\bea
\frac{d}{dR}\bigg[\frac{U {\cal E}
\dot \rho}{\rho^2\sqrt{U R_{D4}^{-3}+\frac{\dot\rho^2}{\rho^2}}}
\bigg]&=&\frac {dU}{d\rho}\frac{{\cal E}}{\sqrt{U R_{D4}^{-3}+
\frac{\dot\rho^2}{\rho^2}}}\bigg(\frac 32 UR_{D4}^{-3}+
\frac{\dot \rho^2}{\rho^2}\bigg)-
\frac{\dot\rho^2U {\cal E}}{\rho^3 \sqrt{U R_{D4}^{-3}
+\frac{\dot\rho^2}{\rho^2}}},\nonumber\\
\eea
where we have introduced the notation
\be
{\cal E}=\sqrt{e^2-(e\omega)^2R^2},
\ee
and dots represent $R$-derivatives.

Given that the NG action (\ref{act}) is invariant under shifts is $X^0$ and
$\theta$, we extract the associated conserved charges, namely 
the energy
\be
E=T_s
\int d R\frac{e\sqrt{(U/R_{D4})^{3/2}+K\dot\rho^2}}{{\cal E}}(U/R_{D4})^{3/4},
\ee
and the angular momentum
\be
J=T_s\int dR\frac{e\omega R^2 \sqrt{(U/R_{D4})^{3/2}+K\dot\rho^2}}
{{\cal E}}(U/R_{D4})^{3/4}.
\ee

We distinguish two region of interest spanned by the open string:

$\bullet$ Region I, characterized by
$\dot\rho \to \infty$,

and 

$\bullet$ Region II, characterized by $\dot\rho\to 0$.

In the limit where the separation between the endpoints of the string
is large, the vertical Region I and the horizontal Region II represent
a good approximation to the shape of the string. However, we should point out 
that the separation distance is not a fixed parameter, rather it is a 
dynamical variable, and we will soon relate it to the external parameters
of the problem: $\omega$ and $\rho_f$. 

Region I can be viewed as extending from $\rho_f$, the location of the 
brane probe, to $\rho_\Lambda$, which is the end of space, at a fixed value
of $R$ which we denote by $\pm R_0$. Region II is the horizontal piece
of the string, with the string extending from $-R_0$ to $R_0$ at fixed
$\rho$.  

Given that we introduced an additional quantity, namely the separation 
between the endpoints $2R_0$, and we decomposed the profile of the string
into two regions, we must re-investigate the way the variational principle is 
satisfied. In (\ref{deltas}) we notice
that there is a surviving boundary term from Region II and a bulk
term from Region I
\bea
&&\int d\tau  \delta R\frac{(U/R_{D4})^{3/2}
\partial_\sigma R \sqrt{(\partial_\tau X^0{}^{2}- R^2 \partial_\tau
\theta{}^2)(U/R_{D4})^{3/2}}}{\sqrt{(U/R_{D4})^{3/2}\partial_\sigma R{}^2 + 
K\partial_\sigma \rho{}^2}}\bigg|_{\sigma=-\alpha}^{\sigma=\alpha}\nonumber\\
&-&\bigg(\int_{\sigma=-\pi/2}^{\sigma=-\alpha}+\int_{\sigma=\alpha}^{\sigma=\pi/2}\bigg)
 d\sigma d\tau\delta R\frac{R\partial_\tau\theta{}^2\sqrt{
(U/R_{D4})^{3/2}((U/R_{D4})^{3/2}\partial_\sigma R{}^2 
+ K\partial_\sigma \rho{}^2)}}{
\sqrt{\partial_\tau X^0{}^{2}- R^2 \partial_\tau
\theta{}^2}}, \nonumber\\
\eea
where Region II corresponds to $\sigma\in (-\alpha,\alpha)$ and Region I
to $\sigma\in(-\pi/2,-\alpha), \sigma\in (\alpha,\pi/2)$.
Substituting the solution to the equations of motion into these terms we
end up with
\bea
\int d\tau \delta R\sqrt{1-\omega^2R_0^2}(U_{\Lambda}/R_{D4})^{3/2}-
\frac{\omega^2R_0}{\sqrt{1-\omega^2 R_0^2}}
\int d\tau \int_{\rho_{\Lambda}}^{\rho_f}d\rho \delta R\frac {U(\rho)}\rho.
\nonumber\\
\eea
The only way we can achieve cancellation is by requiring $\delta R(\rho,\tau)=
\delta R(\tau)$ for $\rho\in(\rho_\Lambda, \rho_f)$ and by enforcing
\be
1-\omega^2 R_0^2=\omega^2 
R_0\frac{1}{(U_\Lambda/R_{D4})^{3/2}}\int_{\rho_{\Lambda}}^{\rho_f}d\rho
\frac {U(\rho)}\rho. \label{cond}
\ee
By recognizing on the rhs the mass of the dynamical quarks
\be
m_q=T_s\int_{\rho_f}^{\rho_\Lambda} d\rho\sqrt{g_{00}g_{\rho\rho}}=
T_s\int_{\rho_f}^{\rho_\Lambda}d\rho\frac{U}{\rho}. \label{mq}
\ee
we found the 
desired relation between the separation distance, angular velocity 
and the mass of the quarks $m_q$ (which is determined as a function of the position of the flavor brane
$\rho_f$). As expected, having the string endpoints 
moving at the speed of light requires $m_q=0$.

Let us analyze in detail the equations of motion.

$\bullet$ Region I:

The equation of motion is satisfied to leading order, as we can see from
\bea
\frac{d}{dR}\bigg(\frac{U{\cal E}}{\rho}\bigg)=\bigg(
\frac{dU}{d\rho}-\frac{U}{\rho}\bigg){\cal E}\frac{\dot \rho}{\rho}.
\eea
Substituting (\ref{uu}) it is easy to check that 
the lhs becomes precisely equal to the rhs.

Assuming an almost rectangular shape, this region can be extended all 
the way to the end of space where the string flattens. 
In terms of $\rho$ the end of space is marked by $\rho_{\Lambda}=
(\frac{1}{2})^{2/3} U_{\Lambda}$.

The contributions to the energy and angular momentum are
\be
E_{I}=T_s\int_{\rho_\Lambda}^{\rho_{f}} d\rho 
\frac{e\sqrt{U^{3/2}(dR/d\rho)^2+K}}{{\cal E}}(U/R_{D4})^{3/4}
=\frac{eT_s}{{\cal E}}
\bigg|_{R=-R_0, R_0}\int d\rho \frac{U}{\rho}
\ee
\be
J_{I}=T_s \frac{e\omega R^2}{{\cal E}}\bigg|_{R=-L/2,L/2}
\int d\rho\frac{U}{\rho}.
\ee
Using that that the mass of the dynamical quarks is given by
(\ref{mq}),
we find the the energy and angular momentum contribution coming from the
vertical regions of the Wilson loop reproduce the energy and angular
momentum of relativistic spinning particles:
\be
E_I=\frac{2m_q}{\sqrt{1-\omega^2 R_0^2}}, \qquad
J_I=\frac{2m_q \omega R_0{}^2}{\sqrt{1-\omega^2 R_0^2}}
\ee

$\bullet$ Region II:

As mentioned before, in this region the string is almost flat in the $\rho$
direction.
We notice that this can happen only for $\rho=\rho_{\Lambda}$, since from the 
equation of motion we find 
\be
\frac{d}{dR}\bigg(\frac{\sqrt U {\cal E}\dot\rho}{\rho^2}\bigg)
=\frac 32\sqrt{U}\frac{dU}{d\rho}R_{D4}^{-3}\,\,{\cal E}.
\ee
In the limit we are interested ($\dot \rho \to 0$), this becomes
\be
\frac{\ddot\rho}{\rho^2}{\cal E}+\frac{\dot {\cal E}\dot \rho}{\rho^2}=
\frac 32 \frac{dU}{d\rho}R_{D4}^{-3}\,\,{\cal E}.
\ee
The lhs vanishes for constant $\rho$ so we require that the rhs vanishes too.
This is indeed the case for  $\rho=\rho_{\Lambda}$, because
 $dU/d\rho|_{\rho_\Lambda}=0$.

The energy and angular momentum contributions of this
region 
\bea
E_{II}&=&T_s\int_{-R_0}^{R_0} 
dR\frac{e\sqrt{U^{3/2}+K\dot \rho^2}}{{\cal E}}(U/R_{D4})^{3/4}\nonumber\\
&=&T_s (U_{\Lambda}/R_{D4})^{3/2}\frac{2}{\omega}\arcsin(\omega R_0)
\\
J_{II}&=&T_s\int_{-R_0}^{R_0} 
dR \,e\omega R^2\frac{e\sqrt{U^{3/2}+K\dot \rho^2}}{{\cal E}}U^{3/4}\nonumber\\
&=&T_s (U_{\Lambda}/R_{D4})^{3/2}
\frac{1}{\omega^2}(\arcsin(\omega R_0)-\omega R_0\sqrt{1-\omega^2
R_0{}^2})
\eea
are those of an open string spinning in 
flat space, but with the string tension rescaled
by $(U_{\Lambda}/R_{D4})^{3/2}$ 
to the value of the gauge theory quark-antiquark 
flux tube string tension $T_g=T_s (U_{\Lambda}/R_{D4})^{3/2}$.

Gathering all the terms, the energy and angular momentum of our Wilson
loop are 
\bea
E&=&T_g\frac{2}{\omega}\arcsin(\omega R_0)+\frac{2m_q}{\sqrt{1-\omega^2 R_0^2}}
\\
J&=&T_g \frac{1}{\omega^2}(\arcsin(\omega R_0)-\omega R_0\sqrt{1-\omega^2
R_0{}^2})+\frac{2m_q \omega R_0{}^2}{\sqrt{1-\omega^2 R_0^2}}.
\eea
Making use of the ``sewing'' condition (\ref{cond}), which
we can now rewrite as
\be
1-\omega^2 R_0^2=\frac{\omega^2 R_0 m_q}{T_g},
\ee
in order to
eliminate the dependence on one of the parameters $\omega$ and 
$R_0$, we find that the energy and angular momentum are given by:
\begin{eqnarray}
\label{trajectory}
E&=&\frac{2T_g}{\omega}\left(\arcsin x +\frac{1}{x}\sqrt{1-x^2}\right),
\nonumber \\
J&=& \frac{T_g}{\omega^2}\left(\arcsin  x + \frac{3}2 x\sqrt{1-x^2}\right),
\end{eqnarray}
where $x =\omega R_0$ is the speed of the endpoints of the string. 

We have thus re-discovered that the stringy meson picture coincides
with the toy model discussed in section 3.
In the limit when $m_q \to 0$, that is when the ends of the string are massless we
recover a linear Regge trajectory when the endpoints of the string
move at the speed of light $x\to 1$. More precisely, keeping the first
two sub-leading terms in $m_q/E$ we find corrections to the linear Regge trajectories of the form 
\be
J=\frac{1}{\pi T_g}E^2\left(1+
\frac{\sqrt{2}}{\sqrt{\pi}}\left(\frac{m_q}{E}\right)^{1/2} -\frac{\pi
  -1 }{\pi}\frac{m_q}{E} +\ldots \right).
\ee
It is interesting to consider the opposite limit, that is, $x\to 0$
\begin{equation}
J=\frac{2 m_q^{1/2}}{T_g}(E-2m_q)^{3/2}
-\frac{11}{12}\frac{1}{m_q^{1/2}T_g}\left(E-2m_q\right)^{5/2}
+\frac{163}{144}\frac{1}{T_g m_q^{3/2}}\left(E-2m_q\right)^{7/2}.
\end{equation}
The above expression is to be understood as correction to the Regge
trajectory for mesons with quark masses $m_q$ in the energy regime where
$E\approx 2 m_q$.

It is also possible to view the corrections in terms of the ratio of
the static energy of the massive of the quarks to the energy of the static string, that is,
$m_q/T_g R_0$. The correspondent expressions are:
\begin{eqnarray}
E&=&\frac{2m_q\sqrt{1+q}}{{q}}\left(\arcsin \frac{1}{\sqrt{1+q}} +\sqrt{q}\right),
\\
J&=& \frac{m_q^2 (1+q)}{T_g q^2}\left(\arcsin \frac{1}{\sqrt{1+q}} 
+ \frac{3}{2}\sqrt{\frac{q}{1+q}}\right),
\end{eqnarray}
where
\be
q=\frac{m_q}{T_g R_0}.
\ee
The limit $q\to 0$ corresponds to the Regge regime whereas the limit
$q\to \infty$
corresponds to the case of non-relativistic motion.

\section{Regge phenomenology}
\label{pheno}

It is remarkable that the Regge trajectory $(J, E^2)$ arising from 
a classical  macroscopic open string spinning
in the non-extremal D4 geometry and ending on a probe D6 brane
coincides with a known phenomenological model. 
Namely, the toy model reviewed in Section \ref{toy}. It is therefore
incumbent upon us to revisit the status of this model in the
phenomenology literature. 

One particular observation is that the Regge
trajectories of mesons are non-linear, with the non-linearity the more 
pronounced the heavier the masses of the constituent quarks.
The trajectories for mesons composed of light quarks
are essentially linear, with a universal value of the slope
$\alpha'\approx 0.85$ GeV$^{-2}$. This class of light-quark mesons
includes, for example,  the trajectories of the $\rho$, $K^{*}$,
$\pi^0$ 
and $\omega$. 
The slope of the Regge trajectories of heavy-quark mesons 
varies along the trajectory,
from a smaller slope for the lightest
states within the trajectory (it is $\approx 0.5$ GeV$^{-2}$
for $c\bar c$), to the same universal slope for the highest
states.

\begin{figure}
\begin{center}
\epsfig{file=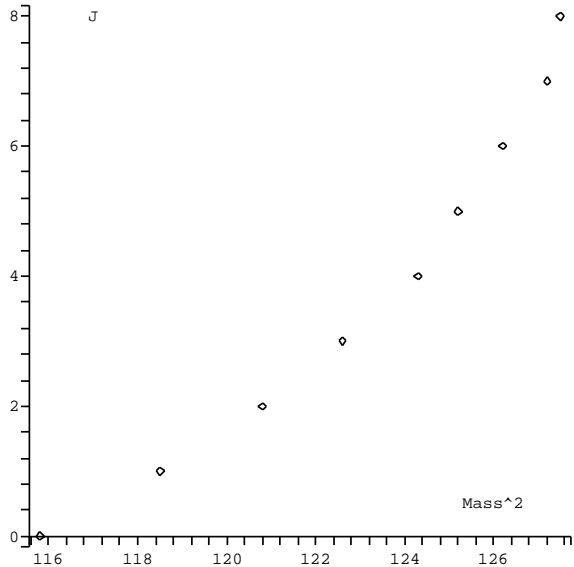,width=9cm}
\caption{ The bottomonium trajectory \cite{bottomdata,bottom}.\label{bb}}
\end{center}
\end{figure}

To exemplify the nonlinearity, we have included the
$b\bar b$  trajectory displayed in Fig. \ref{bb} which is obtained
from \cite{bottom}. The latter paper employs a quark-antiquark
effective potential derived from lattice QCD \cite{bottomdata}. 
Since for $b\bar b$ and $c\bar c$ 
mesons the experimental
data regarding resonances with spin higher than 1 are scarce,
the trajectory in Fig. \ref{bb}
is the result of a lattice analysis. Similar conclusions about the
Regge trajectory can be
drawn for charmonium \cite{cc} and other heavy-quark mesons
\cite{heavyquarks}. 

\begin{figure}
\begin{center}
\epsfig{file=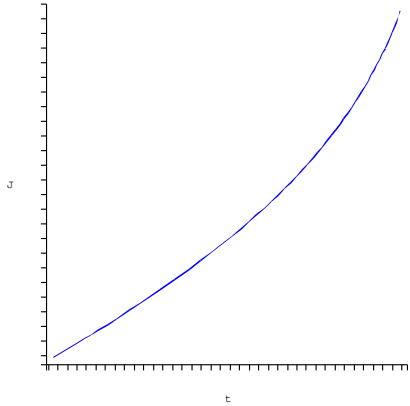,width=6cm}
\caption{A parametric plot of (\ref{trajectory}) for $x\to1$ shows similarity
  with the bottomonium trajectory.\label{rt}}
\end{center}
\end{figure}

Notice that the same flattening of the Regge trajectory for 
the lowest states within the trajectory
can be inferred from plotting $(J, E^2)$ from (\ref{trajectory}), 
as it can be seen from Fig. \ref{rt}.

Thus, the Regge trajectory extracted from the naive toy model 
qualitatively captures various aspects of realistic heavy-quark meson
trajectories. 

The string derivation of the Regge trajectory (\ref{trajectory})
explains why the highest states within any meson 
trajectory end up with the same universal slope. The answer is
beautiful and geometrical: the universal slope
corresponds to the rescaled string tension of the portion of the
string that spins closed to the confining wall, that is, closed to
$U_{\Lambda}$ in figure (\ref{string}). 
This tension $T_g$ is clearly universal and independent on the position 
of the probe branes, hence independent on the flavor of the quarks.

Recently, Wilczek and collaborators \cite{wilczek, wilczek1} have
developed an extremely successful approach to the phenomenology of
mesons and  baryons. Their model includes, as an essential tool,
the mass-loaded version of the Chew-Frautschi model, which is
precisely the effective Regge trajectory  we obtained in this
paper. It would be interesting to approach other aspects of their
model from our fundamental point of view and hopefully provide a {\it
  derivation} from the string model point of view of this
phenomenological model.

\section*{Acknowledgments}

We thank R. Akhoury, O. Aharony and  T. Wang for comments and
suggestions. We are particularly grateful to C. N\'u\~nez and
A. Ramallo for a very detailed reading of the first version of the
manuscript and various comments.  M.K. is grateful to R. Myers and D. Mateos for related
discussions and would like to thank the MCTP for hospitality during the
initial stages of this work. L.P.Z, J.S. and D.V. thank KITP for
hospitality during the late stages of this project, our work at KITP
was supported by an NSF grant. J.S. would also like to thank the
Department of Physics  of the University of
Texas, Austin, where part of this work was done. 
M.K. is supported in part by NSF under grant PHY-0331516 and by DOE under grant
DE-FG02-92ER40706 and a DOE Outstanding Junior Investigator Award. 
L.P.Z. and D.V. are  supported in part by the US Department of Energy
under grant DE-FG02-95ER40899. The work of J.S. was supported in part by the German Israeli Foundation.

\end{document}